\documentstyle[12pt]{article}
\begin{document}
\begin{flushright}
Alberta-Thy-22-97\\
hep-ph/9710275\\
August, 1997\\
\end{flushright}

\begin{center}
{\Large \bf  Inelastic Final-State Interactions and Two-body Hadronic B decays into Single-Isospin channels}\\
\vskip 5mm
A.N.Kamal and C.W.Luo\\
{\em Theoretical Physics Institute and Department of Physics,\\ University of Alberta,
Edmonton, Alberta T6G 2J1, Canada.}\\
\end{center}

\vskip 5mm
\begin{abstract}
The role of inelastic final-state interactions in CP asymmetries and branching ratios  is investigated in  certain chosen single isospin two-body hadronic  B decays. Treating final-state interactions through Pomeron and Regge exchanges, we demonstrate that  inelastic  final state interactions  could lead to sizeable effects on the CP asymmetry.  
\end{abstract}
\par
{\it PACS number(s): 11.30.Er, 13.25.Hw, 13.85.Fb}

\newpage
\begin{center}
{\Large \bf I. Introduction}
\end{center}

It is well-known that CP asymmetries occur in two-body hadronic decays of B meson involving two distinct CKM angles and two differing strong phases\cite{KPS}. The sources of the strong phases are several: perturbative penguin loops,  final-state interaction (fsi) phases involving two different isospins, and  inelastic final-state interactions involving a single isospin state.

\par
In this paper we have studied  the effect of interchannel mixing  on  B decays into two-body single-isospin channels. For reasons to follow,  we have chosen to study the following  decay modes: $B^-\rightarrow  \eta_c \pi^- $, $B^-\rightarrow \eta_cK^- $ and  $B^-\rightarrow \phi K^-$. 
The first of them, $B^-\rightarrow \eta_c\pi^- $,  is a color-suppressed decay into a state with $I=1 $.  At the tree-level, it proceeds through a CKM angle product $V_{cb}V_{cd}^* $. In absence of interchannel mixing, the CP asymmetry for this mode is known to vanish\cite{KPS1}.(Though ref.\cite{KPS1} does not include electromagnetic penguins, their inclusion does not alter this fact.)  The color-favored decay channel $B^-\rightarrow D^0D^- $ with $ I=1 $, also proceeds through a CKM product  $V_{cb}V_{cd}^* $ at the tree-level, and has a nonvanishing CP asymmetry\cite{KPS1}.  An inelastic coupling of $D^0D^- $ channel to $\eta_c\pi^- $ allows a two-step decay $B^-\rightarrow D^0D^-\rightarrow \eta_c\pi^- $ resulting in, as we show later, a significant CP asymmetry in $\eta_c\pi^- $ final state.  
The same can be argued for $B^-\rightarrow \eta_cK^- $,  also a color-suppressed decay proceeding via $V_{cb}V^*_{cs} $ at the tree-level.  It is also known to have a vanishing CP asymmetry\cite{KPS1} in absence of inelastic fsi. However, the interchannel mixing of $\eta_cK^- $ channel with $D^0D^-_s $, a color-favored channel, results in a nonvanishing CP asymmetry in $\eta_cK^- $ mode.  Lastly,
the mode $B^-\rightarrow \phi K^- $ involves $b\rightarrow s\bar{s}s $ decay and proceeds only through a penguin  amplitude involving CKM angle products $V_{ub}V_{us}^* $ and $V_{cb}V_{cs}^* $.  In absence of inelastic fsi, it is known to have a nonzero CP asymmetry. The strong phases here arise from the light quark penguin loops. However,  $\phi K^- $ channel can couple to $ D^{*0}D^-_s $ and $D^0D^{*-}_s $ channels. The decays $B^-\rightarrow  D^{*0}D^-_s $ and $B^-\rightarrow D^0D^{*-}_s $  being Cabibbo-favored, have  branching ratios three orders of magnitude larger than that of the penguin  process $B^-\rightarrow \phi K^- $\cite{KPS1}. We have studied the effect of the inelastic coupling  of the channels  $B^-\rightarrow  D^{*0}D^-_s $ and $D^0D^{*-}_s $ to $B^-\rightarrow \phi K^- $  channel on the branching ratio and CP asymmetry in  the latter mode.

This paper is organized as follows:  In section II, we describe the formalism and  investigate  
the effect of inelastic  fsi  on  the branching ratios and CP asymmetries  in $B^-\rightarrow \eta_c\pi^- $, $B^-\rightarrow \eta_cK^- $ and $B^-\rightarrow \phi K^- $.  The results are discussed in  section III,   .

\begin{center} 
{\Large \bf II.  CP Asymmetry and  Final State Interaction in $B^-\rightarrow \eta_c\pi^- $, $B^-\rightarrow \eta_c K^- $  and $B^-\rightarrow \phi K^- $ }
 \end{center}

\begin{center} 
{\bf A.     Definitions and Formalism}
 \end{center}

\par
The effective Hamiltonian for $b\rightarrow s $ transition (for $b\rightarrow d $, replace $ s $ by $ d $) is given by \cite{DH, F, BJLW}

\begin{equation}
H_{eff}={G_F  \over \sqrt{2} } \sum_{q=u,c}{\left\{  V_{qb} V^*_{qs}[ C_1 O^q_1+C_2 O^q_2+
\sum_{i=3}^{10}{C_i O_i}]  \right\} } .
\end{equation}
The operators in Eq.(1) are the following;
\begin{eqnarray}
O^q_1=(\bar{s}q)_{V-A} (\bar{q}b)_{V-A},   & O^q_2=(\bar{s}_{\alpha}q_{\beta})_{V-A} (\bar{q}_{\beta} b_{\alpha})_{V-A}; \nonumber \\
O_3=(\bar{s}b)_{V-A}\sum_{q^{\prime}} (\bar{q}^{\prime}q^{\prime})_{V-A},  & 
O_4=(\bar{s}_{\alpha} b_{\beta})_{V-A}\sum_{q^{\prime}} (\bar{q}^{\prime}_{\beta} q^{\prime}_{\alpha})_{V-A},     \nonumber  \\
O_5=(\bar{s}b)_{V-A}\sum_{q^{\prime}} {(\bar{q}^{\prime}q^{\prime})_{V+A}}, & 
O_6=(\bar{s}_{\alpha} b_{\beta})_{V-A}\sum_{q^{\prime}}{(\bar{q}^{\prime}_{\beta} q^{\prime}_{\alpha})_{V+A}};        \\
O_7={3 \over 2}(\bar{s}b)_{V-A}\sum_{q^{\prime}} {(e_{q^{\prime}}\bar{q}^{\prime}q^{\prime})_{V+A}}, & 
O_8={3\over 2}(\bar{s}_{\alpha} b_{\beta})_{V-A} \sum_{q^{\prime}}{(e_{q^{\prime}}\bar{q}^{\prime}_{\beta} q^{\prime}_{\alpha})_{V+A}},
\nonumber    \\
O_9={3 \over 2}(\bar{s}b)_{V-A}\sum_{q^{\prime}} {(e_{q^{\prime}}\bar{q}^{\prime}q^{\prime})_{V-A}}, & 
O_{10}={3\over 2}(\bar{s}_{\alpha} b_{\beta})_{V-A} \sum_{q^{\prime}}{(e_{q^{\prime}}\bar{q}^{\prime}_{\beta} q^{\prime}_{\alpha})_{V-A}}.  \nonumber  
\end{eqnarray}
$O_1 $ and $O_2 $ are the Tree Operators, $O_3, ...., O_6 $ are generated by QCD Penguins and $O_7,...., O_{10} $ are generated by Electroweak Penguins.  Here  $V\pm A $ represent $\gamma_{\mu} (1\pm \gamma_5)$, $\alpha$ and $\beta$ are color indices.  $\sum_{q^{\prime}}$ is a sum over the active flavors u,d,s and c quarks.

\par
In the next-to-leading-log calculation one works with effective Wilson coefficients $C^{eff}_i $, rather than the coefficients that appear in (1). The derivation of these effective coefficients is well known \cite{DH, F, BJLW}. We simply quote their values

\begin{eqnarray}
C^{eff}_1=\bar{C}_1,~~  C^{eff}_2=\bar{C}_2, & ~~ C^{eff}_3=\bar{C}_3 - P_s / N_c,  & C^{eff}_4=\bar{C}_4 + P_s,  \nonumber   \\
C^{eff}_5=\bar{C}_5 - P_s/N_c,~~~~ &  C^{eff}_6=\bar{C}_6 +P_s, &  C^{eff}_7=\bar{C}_7 +P_e, 
 \nonumber   \\
 C^{eff}_8=\bar{C}_8, ~~~~~~~~~~~~~~ & C^{eff}_9=\bar{C}_9 +P_e, &  C^{eff}_{10}=\bar{C}_{10},
\end{eqnarray}
with\cite{DH}
\begin{eqnarray}
& \bar{C}_1=1.1502, ~\bar{C}_2=-0.3125, ~\bar{C}_3=0.0174, ~\bar{C}_4=-0.0373, ~\bar{C}_5=0.0104,~ \bar{C}_6=-0.0459, & \nonumber   \\
& \bar{C}_7=-1.050\times 10^{-5}, ~~\bar{C}_8=3.839\times 10^{-4}, ~~\bar{C}_9=-0.0101, 
~~\bar{C}_{10}=1.959\times 10^{-3},~~~~~~&
\end{eqnarray}
and 
\begin{eqnarray}
P_s &= &{\alpha_s (\mu) \over  8\pi} C_1 (\mu) [{10 \over 9}+\frac{2}{3}\ln{\frac{m_q^2}{\mu^2}}-G(m_q, \mu, q^2)], \\
P_e &= &{\alpha_{em} (\mu) \over  3\pi} [C_2 (\mu)+{C_1(\mu) \over N_c}] [{10 \over 9}+\frac{2}{3}\log{\frac{m_q^2}{\mu^2}}-G(m_q, \mu, q^2)],
\end{eqnarray}
where
\begin{equation}
G(m_q, \mu, q^2)= -4\int_{0}^{1}{dx x (1-x) \ln{[1-x (1-x) {q^2 \over  m^2_q}}]},
\end{equation}
$q^2 $ is the   momentum carried by the gluon or the photon in the penguin diagram and 
$m_q$ the mass of the quark q in the penguin loop.   For  $q^2> 4 m^2_q$,  $G(m_q, \mu, q^2)$ becomes complex giving rise to   strong perturbative phases through  $P_s $ and $P_e $. 
The parameters we employ are:
\begin{eqnarray}
& m_u=5 MeV  ,   m_s=175MeV   ,   m_c=1.35GeV , m_b=5.0GeV,    & \nonumber  \\
& CKM angles:  A=0.81, \lambda=0.22,  (\rho, \eta )=(-0.20, 0.45)~ and~ (0.30, 0.42),  & \nonumber  \\
&  f_D=200MeV  , f_{D_s}=f_{D^*_s}=f_{\eta_c}=300 MeV,   f_{\phi}=233MeV.     &   \\
\end{eqnarray}
\par

Consider now each one of the decays $B^-\rightarrow \eta_c\pi^-,   \eta_c K^- $ and $\phi K^- $ in absence of inelastic fsi. In the factorization approximation,  which we adopt, the decay amplitudes are:
\begin{eqnarray}
A(B^-\rightarrow \eta_c \pi^- )&= &\frac{G_F}{\sqrt{2}}\{ V_{cb}V^*_{cd} a_2-V_{tb}V_{td}^* (a_3-a_5-a_7+a_9)\}<\eta_c|(\bar{c}c)_{V-A}|0><\pi^-|(\bar{d}b)_{V-A}|B^->\nonumber \\
& & \\
A(B^-\rightarrow \eta_c K^- )&= &\frac{G_F}{\sqrt{2}}\{ V_{cb}V^*_{cs} a_2-V_{tb}V_{ts}^* (a_3-a_5-a_7+a_9)\}<\eta_c|(\bar{c}c)_{V-A}|0><K^-|(\bar{s}b)_{V-A}|B^->\nonumber \\
& & \\
A(B^-\rightarrow \phi K^- ) &=&-\frac{G_F}{\sqrt{2}}V_{tb}V^*_{ts} \{ (1+\frac{1}{N_c} )C^{eff}_3
+ (1+\frac{1}{N_c} )C^{eff}_4+a_5-a_7/2-\frac{1}{2} (1+\frac{1}{N_c} )C^{eff}_9\nonumber \\ 
& &  -\frac{1}{2} (1+\frac{1}{N_c} )C^{eff}_{10} \}<\phi|(\bar{s}s )_{V-A} |0><K^-|(\bar{s}b)_{V-A}|B^->
\end{eqnarray}
where we have used the unitarity relation $\sum_{q=u, c} V_{qb}V^*_{qs(d)}=-V_{tb}V^*_{ts(d)}  $ and defined 
\begin{eqnarray}
a_{2i}=C^{eff}_{2i}+\frac{1}{N_c}C^{eff}_{2i-1}, \nonumber \\
a_{2i-1}=C^{eff}_{2i-1}+\frac{1}{N_c}C^{eff}_{2i},
\end{eqnarray}
with  i an integer.  The strong phases appear through $P_s $ and $P_e $ defined in (5) and (6).
However,  in  (10 ) and ( 11) , odd coefficients $a_3, a_5, a_7 $ and $a_9 $ involve  such combinations of $C^{eff}_i $  as to cancel the effect of $P_s $ and $P_e $. Thus,  strong phases do not
appear in (10 ) and (11 )  but they do in (12 ). Hence,  CP asymmetry vanishes for $B^-\rightarrow \eta_c\pi^- $ and $\eta_c K^- $ but is nonzero for $B^-\rightarrow \phi K^- $. This is true even if the electromagnetic penguins were ignored as in ref\cite{KPS1}.

Let us now consider the decay channel $B^-\rightarrow D^0D^- $. The decay amplitude is given by,
\begin{equation}
A(B^-\rightarrow D^0D^- )=\frac{G_F}{\sqrt{2}}\{ V_{cb}V_{cd}^*a_1-V_{tb}V^*_{td}(a_4+(a_6+a_8)R_1+a_{10})\} <D^-|(\bar{d}c)_{V-A}|0><D^0|(\bar{c}b)_{V-A}|B^->
\end{equation}
where
\begin{equation}
R_1=\frac{2m^2_D}{(m_b-m_c)(m_c+m_d)}.
\end{equation}
Note that the above decay is color-favored (tree diagram being proportional to $a_1 $ ) and that strong phases do not cancel in the even coefficients $a_4, a_6, a_8 $ and $ a_{10} $.  Hence CP asymmetry in $B^-\rightarrow D^0D^- $ is nonvanishing\cite{KPS1}.
Throughout our calculations, we have used the formfactors from Bauer, Stech and Wirbel\cite{BSW}.
\par
In the following section we discuss in detail the mixing of $D^0D^- $ and $\eta_c\pi^- $ channels through inelstic fsi.

\begin{center}
{\bf  B.    Inelastic mixing of $D^0D^- $ and $\eta_c\pi^- $ channels. }
\end{center}

Inelastic fsi have been discussed in the past\cite{KSS, BRS,WS, Kamal, KL} in the context of the K-matrix formalism. The desirable feature of this method is that unitarity of the S-matrix  is ensured. The difficulty lies in the proliferation of  K-matrix parameters, mostly unknown, with the number of channels.  Moreover, in two-channel problems, the second channel (the inelastic channel ) is assumed to reflect (through unitarity) all the inelastic channels.   This is an oversimplification of reality.  In ref.\cite{BRS},  the coupling of $\eta_c\pi^- $  channel to 
$\eta\pi^- $ and $\eta^{\prime}\pi^- $ is discussed in K-matrix formalism.   We comment on this work in Section III.

In the calculations we present below, we make no effort to enforce two-channel unitarity.  Rather, we couple the decay channel $\eta_c\pi^- $ to $D^0D^- $  using a Regge-exchange model.   The model coupling constants are  related to known coupling constants by approximations we explain in the text.  The advantage of this procedure, in contrast to the K-matrix approach\cite{KSS,BRS,WS,Kamal, KL}, is that the scattering parameters are determined more realistically. The shortcoming  is that the relevant elements of the  S-matrix  being  completely determined,  the  S-matrix  itself does not satisfy two-channel unitarity. Yet, we think it is more realistic to treat the effect of inelastic channels  one channel at a time rather than enforce two-channel unitarity on what is in fact a multi-channel problem.
\par
We begin by establishing certain key equations for an arbitrary number of two-body channels. An $n\times n $ S-matrix  for s-wave scattering satisfying unitarity can be written as \cite{Newton}
\begin{equation}
\bf{S}=(1+i\bf{k}^{\frac{1}{2}}\bf{K}\bf{k}^{\frac{1}{2}}) (1-i\bf{k}^{\frac{1}{2}}\bf{K}\bf{k}^{\frac{1}{2}})^{-1},
\end{equation}
where $\bf{k} $ is a digonal momentum matrix and $\bf{K} $ a real-symmetric matrix with $n(n+1)/2 $ real parameters.  The decay amplitudes for the B meson into n two-body channels are inelastically coupled through
\begin{equation}
\bf{A}=(1-i\bf{k}^{\frac{1}{2}}\bf{K}\bf{k}^{\frac{1}{2}})^{-1}\bf{A^{(0)}},
\end{equation}
where $\bf{A}^{(0)} $ is a column of uncoupled amplitudes. The coupled (unitarized ) amplitudes are assembled in the column $\bf{A} $.
From (16 ) and (17)  it is easily shown that \cite{KL}
\begin{equation}
\bf{A}=\frac{1+\bf{S}}{2}\bf{A}^{(0)}.
\end{equation}

Let us label channels $\eta_c\pi^- $ and $D^0D^- $ as channels 1 and 2 respectively.   $A_1^{(0)} $ and $A_2^{(0)} $ are given by (10 ) and (14) respectively. In order to calculate  the effect  of channel 2 on channel 1 and vice-versa, we need to calculate  the elements  $S_{11} $, $ S_{12} $ and $S_{22} $ of the S-matrix .  We describe  their evaluation  in the following.
\par
	We assume that Pomeron exchange dominates  elastic scattering.  The scatterings $\eta_c\pi^- \rightarrow \eta_c\pi^- $  and $D^0D^-\rightarrow D^0D^- $ are then represented by amplitudes of the form\cite{Collins}
\begin{equation}
P(s, t)=\beta(t) (\frac{s}{s_0})^{\alpha_P(t)} e^{i\pi\alpha_P(t)/2}
\end{equation}
where $\sqrt{s_0} $ is an energy scale and the Pomeron trajectory  is parameterized by 
\begin{equation}
\alpha_P(t)=1.08+0.25t.
\end{equation}
The momentum transfer t is expressed in $GeV^2 $ in the above.  The Pomeron coupling strength,  $\beta(t) $ , is assumed\cite{Zheng, NP}  to have  a t-dependence of the form $\beta(t)=\beta(0) e^{2.8t} $. In the additive quark model, $\beta(0)=4\beta(cu) $ for $\eta_c\pi^-\rightarrow \eta_c\pi^- $, and $\beta(0)=2\beta(cu)+\beta(uu)+\beta(cc) $ for $D^0D^-\rightarrow D^0 D^- $ scattering. The residue $\beta(uu) $ can be extracted from high energy $pp $ and $\pi p $ scattering data yielding \cite{Zheng, NP} $\beta(uu) \approx 6.5 $. No experimental  information exists for the determination of $\beta(cu) $. We make the theoretical ansatz\cite{NP}:  $\beta(cu)\approx \frac{1}{10}\beta(uu) $, and  assume $\beta(cc) $ to be negligibly small\cite{NP}

The inelastic scattering $\eta_c\pi^- \rightarrow D^0D^- $ is mediated by $D^{*0} $ Regge-exchange in the t channel (defined by $t=(P_{D^0}-P_{\eta_c})^2 $ ). The amplitude is of the form \cite{BH}
\begin{equation}
R(s, t)=\beta_R(t) \frac{1-exp[-i\pi \alpha(t)]}{sin[\pi\alpha(t)]}(\frac{s}{s_0})^{\alpha(t)}.
\end{equation}
For $\beta_R(t) $ we adopt a t  dependence\cite{Collins, BH, VK}
\begin{equation}
\beta_R(t)=\frac{\beta_R(0)}{\Gamma[\alpha(t)]}.
\end{equation}
The fact that $\Gamma(z) $ has simple poles at $z=0, -1, -2, ..., $ ensures that the Regge amplitude (21) does not develop nonsense poles at $\alpha(t)=0, -1, -2, ...$. We also note that  in addition to $R(s, t) $  of (21), there is a u-channel exchange amplitude $R(s, u) $ generated by a charged $D^* $ exchange.

Generally, $s_0$  is expected to be process-dependent. For light mesons and baryons it has been taken\cite{Collins} as $s_0=\frac{1}{\alpha^{\prime}}\approx 1 GeV^{2} $. However, for heavy mesons and baryons, the scale $s_0 $ must reflect somehow a higher threshold for the scattering processes. Based on the work of \cite{VK}, it is argued in \cite{BH} that for $\pi D $ scattering mediated by $\rho $-trajectory, $s^{\pi D}_0\approx \frac{2}{\alpha^{\prime}_R} $. We assume this value for $\eta_c\pi^-\rightarrow D^0D^- $ scattering amplitude.

We determine $\beta_R(0) $ by taking the limit $\alpha(t)\rightarrow 1 $ ($D^* $ pole ) in (21 ) and comparing it with the perturbative t-channel $D^* $-pole diagram.  For the latter we assume a $VPP $ vertex of form:
\begin{equation}
L^{VPP}_{int}=f_{ijk}g_{VPP}V^i_{\mu}P^j\partial^{\mu}P^k,
\end{equation}
where i, j and k are SU(4) labels and $f_{ijk} $ the antisymmetric symbol.
\par
The perturbative t-channel $D^*$-exchange graph yields,
\begin{equation}
R(s\rightarrow \infty,  t\sim m^2_{D^*})=g^2_{VPP} \frac{s}{t-m_{D^*}^2}.
\end{equation}
Comparing the limiting case ($t\rightarrow m^2_{D^*}, \alpha(m^2_{D^*})=1) $ of  (21 ) with (24 ) results in
\begin{equation}
\beta_R(0)=\pi g^2_{VPP}.
\end{equation}
SU(4) symmetry allows us to determine $g_{VPP} $ from $\rho\rightarrow \pi\pi $ and $K^*\rightarrow K\pi $ decays\cite{TK},
\begin{equation}
\frac{g^2_{VPP}}{4\pi}\approx 3.0.
\end{equation}
Heavy Quark effective Theory(HQET)\cite{Yan,SS,CDBGFN,KX} could  also have been used to determine $g_{D^*D\pi} $ if the rate $\Gamma(D^*\rightarrow D\pi) $ were known. In absence of this information, authors of ref.\cite{Yan} fix this coupling  by constraining it to yield the axial coupling of the nucleon, $g_A\approx 1.25 $. This results in
\begin{equation}
g_{D^{*-}D^0\pi^-}=g_{D^{*0}D^+\pi^-}=g_{D^{*0}_sD^+K^-}\approx \frac{3\sqrt{m_Dm_{D^*}}}{4f_{\pi}}~~~to~~~~\frac{\sqrt{m_Dm_{D^*}}}{f_{\pi}},
\end{equation}
where $f_{\pi}=131MeV $. This is a much larger coupling constant  than that implied in \cite{TK}, 
 resulting in  $\Gamma(D^{*-}\rightarrow D^0\pi^-)=(100-180)KeV $\cite{Yan}.  In contrast, the SU(4) symmetry scheme of ref\cite{TK} obtains    $\Gamma(D^{*-}\rightarrow D^0\pi^-)= 16KeV $ using $g_{VPP} $ given in (26 ). In our calculations we use the SU(4) symmetry coupling given by (23) and (26) only.

From the scattering amplitudes we project out the elements of the S-wave scattering matrix
\begin{equation}
S_{ij}=\delta_{ij}+\frac{i}{8\pi \sqrt{\lambda_i\lambda_j}}\int^{t_{max}}_{t_{min}}dt T(s, t),
\end{equation}
where $T(s, t) $ is the total amplitude, $\lambda_i $ and $\lambda_j $ are the usual triangle functions $\lambda(x, y, z)=(x^2+y^2+z^2-2xy-2xz-2yz)^{1/2} $ for channels i and j respectively
and $t_{max} $, $t_{min} $ are the limits of the momentum transfer.  We also took into account the u-channel charged $D^*$-exchange in calculating the S-matrix elements. The resulting S-matrix elements (channel 1 = $\eta_c\pi^- $, channel 2 =$D^0D^- $) are,
\begin{equation}
S=\pmatrix{
0.946-0.93\times 10^{-3}i &   0.19\times 10^{-2}+0.068i  &  .....\cr
0.19\times 10^{-2}+0.068i  &  0.843-0.27\times 10^{-2}i  & .....\cr
  .    & . & ..... \cr
 }.
\end{equation}
Clearly, two-channel unitarity is not satisfied but the S-matrix elements $S_{ij} (i, j=1,2) $ are
completely determined. Calculation of the unitarized decay amplitudes proceeds by using $A^{(0)}_i $ from (10 )  and (14), and $\bf{S} $  from ( 28) in (18). The calculation of the branching ratios and CP asymmetry is then straight forward.  We have chosen to perform the calculation for $N_c=3 $ and $N_c=2.4$. The latter choice, suggested in \cite{KP}, could be interpreted to reflect nonfactorization effects. The results are shown in Tables 1 and 2.
 
\par
We note from  these Tables  that  the induced CP asymmetry in $\eta_c \pi $ channel is large;  in fact,  as large as in  channel $D^0D^- $ to which it  is coupled. The CP asymmetry in $\eta_c\pi^- $ channel, however, depends almost linearly on $g^2_{VPP} $. Thus,  increasing (decreasing) $g_{VPP} $ by a factor of 2 results in an increase (decrease)  of CP asymmetry by approximately a factor of four.  We defer the discussion of the results to Section III. 

\begin{center}
{\bf   C.    Inelastic Mixing of $D^0D^-_s $ and $\eta_cK^- $ channels
}
\end{center}
In absence of interchannel coupling, the decay amplitude for $B^-\rightarrow D^0D^-_s $ is given by
\begin{equation}
A(B^-\rightarrow D^0D^- )=\frac{G_F}{\sqrt{2}}\{ V_{cb}V_{cs}^*a_1-V_{tb}V^*_{ts}(a_4+(a_6+a_8)R_2+a_{10})\} <D_s^-|(\bar{s}c)_{V-A}|0><D^0|(\bar{c}b)_{V-A}|B^->,
\end{equation}
where
\begin{equation}
R_2=\frac{2m^2_{D_s}}{(m_b-m_c)(m_c+m_s)}.
\end{equation}

Let us label $\eta_cK^- $ and $D^0D^-_s $ as channels 1 and 2 respectively. The pomeron-mediated elastic scattering now involves coupling constants  $2\beta(cu)+2\beta(cs) $ for $\eta_cK^- $ channel and $\beta(cc)+\beta(cu)+\beta(cs)+\beta(us) $ for $D^0D^-_s $ channel. For $\beta(cs) $ we use the ansatz: $\beta(cs)\approx \frac{1}{10}\beta(us)\approx \frac{1}{15}\beta(uu) $. The Pomeron amplitude is then given as in (19).

The inelastic  scattering $\eta_cK^-\rightarrow D^0D^-_s $ is mediated by $D^{*0} $-exchange in the t channel ($t=(P_{D^0}-P_{\eta_c})^2 $), and by $D^*_s $-exchange in the u channel. The calculation  of  the effect of inelastic coupling of $\eta_c K^- $ and $D^0 D^-_s $ channels parallels that of $\eta_c\pi^- $ and $D^0D^- $ channels described in the previous section. 
The resulting S-matrix ($\eta_cK^- $ = channel 1, $ D^0 D^-_s $ = channel  2 ) is 
\begin{equation}
S=\pmatrix{
0.954-0.8\times 10^{-3}i &  0.85\times 10^{-3}+0.068i  &  .....\cr
0.85\times 10^{-3}+0.068i  &  0.888-0.19\times 10^{-2}i  & .....\cr
  .    & . & ..... \cr
 }.
\end{equation}
The resulting branching ratios and CP asymmetries are shown in Tables 1 and 2. Again, we notice that CP asymmetry induced in channel $\eta_cK^- $ is comparable to that in channel $D^0D^-_s $. Further discussion of the results is deferred to section III.

\begin{center}
{\bf D.Inelastic coupling of $\phi K^- $ to $D^{*0}D^-_s $ and $D^{*-}_sD^0 $ channels.
}
\end{center}
In absence of inelastic fsi, the decay amplitudes for $B^-\rightarrow D^{*0}D^-_s $ and $D^{*-}_s D^0 $ are:
\begin{eqnarray}
A(B^-\rightarrow D^{*0}D^-_s) &= &\frac{G_F}{\sqrt{2}}\{ V_{cb}V_{cs}^*a_1-V_{tb}V^*_{ts}(a_4-(a_6+a_8)R_3+a_{10})\} <D_s^-|(\bar{s}c)_{V-A}|0><D^{*0}|(\bar{c}b)_{V-A}|B^->,  \\
A(B^-\rightarrow D^0D^{*-}_s) &= &\frac{G_F}{\sqrt{2}}\{ V_{cb}V_{cs}^*a_1-V_{tb}V^*_{ts}(a_4+a_{10})\} <D_s^{*-}|(\bar{s}c)_{V-A}|0><D^0|(\bar{c}b)_{V-A}|B^->,
\end{eqnarray}
where
\begin{equation}
R_3=\frac{2m^2_{D_s}}{(m_s+m_c)(m_c+m_b)}.
\end{equation}

Inelastic fsi couple the amplitude for $B^-\rightarrow \phi K^- $, eq.(12), to the amplitudes in (33) and (34).  The calculation of the S-matrix elements ($\phi K^- $= channel 1, $D^-_sD^{0*} $= channel 2,  $D^{*-}_sD^0 $= channel 3) is  considerably simplified in the B  rest-frame. This is because in this frame the vector meson can only have longitudinal helicity in a $B\rightarrow VP $ decay. Because of this fact, the Pomeron and Regge amplitudes involving only helicity  0$\rightarrow $helicity 0 transition are the same as in spin-less scattering. The Pomeron amplitude for elastic scattering is given by (19) with $\beta(0)=2\beta(su)+2\beta(ss) $ for $\phi K^- $ elastic scattering, and $\beta(0)=\beta(us)+\beta(cu)+\beta(cs)+\beta(cc) $ for $D^-_sD^{*0} $ and $D^{*-}_sD^0 $ elastic scatterings. We assume $\beta(ss)\approx \frac{2}{3}\beta(su) $.

\par
Inelastic scatterings,  $\phi K^-\rightarrow D^-_sD^{*0} $ and $D^{*-}_sD^0 $,   are mediated by $D^*_s $ exchange in the t channel ($t=(P_{D_s}-P_{\phi})^2 $ and $ (P_{D^*_s}-P_{\phi})^2 $ respectively). There are no u channel exchanges. $D_s$-trajectory , being a lower-lying trajectory, makes a smaller contribution and we neglect it. The Regge-amplitude is assumed to be of the form given in (21). In order to determine the coupling $\beta(0) $,  we equate the limiting form of (21) for $t\rightarrow m^2_{D^*_s} $ with the perturbative  expressions for  $\phi K^-\rightarrow D^-_sD^{*0} $ and $D^{*-}_sD^0 $ with $D^*_s $ exchange.   We adopt the following definitions,

\begin{eqnarray}
D^*_s -trajectory: & \alpha_{D^*_s}\approx -1.23+0.5t, \\
VVP-vertex:            &  L_{int}(VVP)=g_{VVP}d_{ijk}\epsilon_{\mu\nu\rho\sigma}\partial_{\mu}V^i_{\nu}\partial_{\rho}V^j_{\sigma}P^k ,       \\
VVV-vertex: &  L_{int}(VVV)=g_{VVV}f_{ijk}(\partial_{\mu}V^i_{\nu}-\partial_{\nu}V^i_{\mu})V^j_{\mu}V^k_{\nu}, 
\end{eqnarray}
where $d_{ijk} $ and $f_{ijk} $ are SU(4) indices. The coupling $g_{VVP} $ has dimension $(mass)^{-1} $ while  $g_{VVV} $ is dimensionless.  $D^*_s $ trajectory is assumed to be parallel to $D^* $ trajectory with a slope as in\cite{Zheng}. 

The evaluation of perturbative $D^*_s$-exchange diagram was done numerically. The calculation was made  simpler by the fact that the vector particles could only be longitudinally polarized in B rest-frame. For large s we obtain for the digrams shown in Figs.1 and 2,
\begin{eqnarray}
T^{Fig.1}(s\rightarrow\infty, t\sim m_{D^*_s}^2)=  0.7GeV^2g_{D^*D^*_sK}g_{D_sD^*_s\phi}\frac{s}{t-m_{D^*_s}^2}, \nonumber \\
T^{Fig.2}(s\rightarrow \infty, t\sim m_{D^*_s}^2)= 0.5g_{DD^*_sK}g_{D^*_sD^*_s\phi}  \frac{s}{t-m_{D^*_s}^2}.    
\end{eqnarray}

The corresponding Regge-exchange amplitude yields
\begin{equation}
T(s\rightarrow\infty, t\sim m_{D^*_s}^2)=\frac{\beta(0)}{\pi }\frac{s}{t-m_{D_s^*}^2},
\end{equation}
where we have used $s_0=2/\alpha^{\prime} $.
\par
A comparison of (40) with  (39) yields $\beta(0) $. The coupling constants in (39) are determined as follows:  In HQET\cite{Yan}, where light pseudoscalar mesons are introduced as nonlinear realization of $SU(3)_L\times SU(3)_R $, one obtains ($f_{\pi}=131MeV $,  and we are using the parameter\cite{Yan} f=-1.5),
\begin{equation}
g_{D^{*0}D^{*-}_sK^- }\approx \frac{3}{4f_{\pi}}.
\end{equation}
Light vector and axial-vector mesons can also be introduced in HQET\cite{SS,CDBGFN,KX} allowing a $D^*_sD_s\phi $ coupling. However,  HQET does not  by itself permit an evaluation of  this coupling constant.   Use of vector dominance in the  radiative decays of light mesons determines \cite{BJE} $g_{VVP}\approx 6GeV^{-1} $. Using SU(4) symmetry, one obtains  
$g_{D^{*-}_sD^-_s\phi}\approx 6/\sqrt{2} GeV^{-1} $ which is a little lower than the value given in (41). For want of a better choice we assume $g_{D^{*-}_sD^-\phi}=g_{D^{*0}D^{*-}_sK^-} $.

HQET\cite{SS,CDBGFN,KX} allows us to calculate the VVV coupling $g_{D^*_sD^*_s\phi} $ provided an assumption is made as to how the flavor singlet and the flavor octet of the light vector meson couple to $D^*_sD^*_s $. In nonet symmetry, we obtain
\begin{eqnarray}
g_{D^*_sD^*_s\omega} =0  ~~~~~~~~~~~~~~~~~~~~~~~~~~\\
g_{D^*_sD^*_s\phi}  = g_{VVV} =g ~( ~of ~ [22] ~)\approx 4.3
\end{eqnarray}
As for the VPP coupling, we adopt the value in (26).

To calculate the effect of channels $D^{*0}D^-_s $(channel 2) and  $D^{*-}_sD^0 $(channel 3) 
on $\phi K^- $(channel 1), we need the elements $S_{11}, S_{12}, S_{13} $ of the S-matrix . The decay amplitude for $B^-\rightarrow \phi K^- $ is then given by,
\begin{equation}
A(B^-\rightarrow \phi K^-)  =  \frac{1+S_{11}}{2}A^{(0)}(B^-\rightarrow \phi K^-)+\frac{S_{12}}{2} A^{(0)}(B^-\rightarrow D^{*0}D^-_s)+\frac{S_{13}}{2} A^{(0)}(B^-\rightarrow D^0D^{*-}_s), 
\end{equation}
The relevant elements of the S-matrix are calculated to be,
\begin{eqnarray}
S_{11} &=& 0.784- 0.37\times 10^{-2} i \nonumber \\
S_{12} &= &-0.53\times 10^{-3} + 0.12\times 10^{-2} i \nonumber  \\
S_{13}& =& -0.33\times 10^{-3} + 0.70\times 10^{-3} i
\end{eqnarray}

Since $B^-\rightarrow \phi K^- $ is a penguin mediated process, its branching ratio in absence of inelastic coupling is small $(\sim 10^{-6}~to~10^{-5} )$\cite{KPS1}. In contrast, the inelastic channels $D^{*0}D^-_s$ and $D^0D^{*-}_s $ to which it couples are Cabibbo-favored and have branching ratios of the order of $10^{-3} $ to $10^{-2} $\cite{KPS1}. Thus, whereas the  two channels $D^{*0}D^-_s $ and $D^0D^{*-}_s $ can influence the branching ratios and CP asymmetries in $\phi K^- $, we do not expect the $\phi K^- $ channel to significantly effect the channels $D^{*0}D^-_s$ and $D^0D^{*-} $.  Further, the inelastic  coupling of $D^{*0}D^-_s $ channel to $D^0D^{*-}_s $ occurs via the exchange of $(c\bar{c})$ mesonic trajectories, $\eta_c $ and $\psi $.  As both of these trajectories are low-lying with large and negative intercepts  $\alpha(0) $ , their contribution to the inelastic scattering $D^{*0}D^-_s\rightarrow D^0D^{*-}_s $ is expected to be highly suppressed. We, therefore, do not expect one of these channels to effect the other significantly either.
For these reasons we have displayed in Tables 1 and 2 only the effect on the channel $\phi K^- $. The other two channels, $D^{*0}D^-_s $ and $D^0D^{*-}_s $, are left largely unaffected by fsi.

Tables 1 and 2 show that though the effect of inelastic fsi on the branching ratio for $\phi K^- $ is small (due to the small size of $S_{12} $ and $S_{13} $ in (45)), the effect on the CP asymmetry is significant. The results are discussed in the following section.

\begin{center}
{\Large \bf III. Results and Discussion}
\end{center}

In absence of inelastic fsi,  CP asymmetries in $B^-\rightarrow \eta_c\pi^- $  and $\eta_c K^- $ channels vanish\cite{KPS1}. We have shown  that an inelastic coupling of $\eta_c\pi^- $ channel to  $D^0D^- $ and that of $\eta_cK^- $ to $ D^0D^-_s $, leads to substantial CP  asymmetries in $B^-\rightarrow \eta_c\pi^- $ and $\eta_cK^- $ decays.  We have used $N_c=3 $ and 2.4 and two sets of values for $(\rho, \eta) $. The CP asymmetries depend sensitively on the coupling constant $g_{VPP} $ and the energy-scale parameter $s_0$. An increase (decrease) of $g_{VPP} $ by a factor of two increases (decreases) the CP asymmetry by roughly  a factor of 4.  Similarly, increasing $s_0 $ from $1 GeV^2$ to $2/\alpha^{\prime}_R $ enhances the CP  asymmetry by raising the values of the off-diagonal elements of the S-matrix.  In addition, the calculated CP  asymmetry will also depend on the effective $q^2$ employed.  We have used $q^2=m_b^2/2 $.

Ref.\cite{BRS} discusses a three-channel problem, $B^-\rightarrow \eta\pi^-, \eta^{\prime}\pi^- $ and $\eta_c\pi^- $, in the K-matrix formalism and demonstrates that a CP asymmetry of the order of $\sim 1\% $ can be generated in $\eta_c\pi^- $ channel through inelastic fsi. Our work  differs from \cite{BRS} in several respects. Most importantly, we couple $\eta_c\pi^- $, a color-suppressed channel, to $D^0D^- $, a color-favored channel. The channels, $\eta\pi^- $ and $\eta^{\prime}\pi^- $, invoked in \cite{BRS} are both color-suppressed, and thus, not expected to be as important as $D^0D^- $. In addition, we do not require the S-matrix to be unitary at two or three channel level. Another  important difference between our work and that of \cite{BRS} lies in how the strong interaction fsi parameters are determined. We use Pomeron and Regge phenomenology, presumably applicable at $\sqrt{s}\sim m_B $, while \cite{BRS} determines the K-matrix elements through low-energy phenomenology. For example, the diagonal  and off-diagonal elements of the K-matrix are evaluated  using a contact $\phi^4$-interaction which allows for only S-wave scattering. This is  expected  to be a reasonable approximation at  threshold but hardly likely to hold at $\sqrt{s}\sim m_B $.  Despite these differences, we emphasize that the important conclusion of \cite{BRS} was that a significant CP asymmetry in $\eta_c\pi^- $ could be generated by coupling it to $\eta\pi^- $ and $\eta^{\prime}\pi^- $ channels. However, the fact that they also found asymmetries of the order of $10\% $ and $20\% $  in $\eta\pi^- $ and $\eta^{\prime}\pi^- $ channels has little to do with inelastic fsi;  asymmetries of this magnitude are generated in these channels in absence of inelastic fsi\cite{KPS1}.

\par
We have found that  the CP asymmetry in $B^-\rightarrow \phi K^- $(a penguin driven process) is significantly effected by a coupling to Cabibbo-favored channels  $D^{*0}D^-_s $ and $D^0D^{*-}_s $. Due to the smallness of the off-diagonal elements of the S-matrix , the effect on the branching ratio is not  as large as on CP asymmetry. Again, the CP asymmetry depends sensitively on the coupling constants and the value of $s_0 $.

This work was partly supported by a research grant from the Natural Sciences and Engineering Research Council of Canada to A.N.Kamal.

\newpage
~~~~~~~~

\begin{small}
\begin{table}
\begin{center}
\caption{Branching  ratios and CP asymmetries  with $q^2=m^2_b/2 $  and  $N_c=3 $ ( entries in bracket correspond to the uncoupled case) }
\begin{tabular}{|c|lr|lr|}
\hline
&& \\
CKM  Matrix    &  $\rho=0.30, $ &  $    \eta=0.42  $    &   $ \rho=-0.20,  $   &   $   \eta=0.45 $   \\
\hline
Decay channel  &   BR                   &  $         a_{CP}(\%)       $  &     BR            &         $ a_{CP}(\%)       $     \\
\hline
$B^-\rightarrow \eta_c\pi^- $        &  $2.85\times 10^{-6}  $&               3.97   &  $2.87\times 10^{-6}  $ &   
4.24    \\
  	                                                  & $(2.32\times 10^{-6}) $  &             (0.00 )   &  $ (2.33\times 10^{-6} ) $      &  (0.00 ) \\

$B^-\rightarrow D^0D^- $              &  $3.55\times 10^{-4} $      & 3.94      & $3.22\times 10^{-4} $   &   4.69  \\
						&  $ (4.18\times 10^{-4}) $  & (3.97)   &  $(3.80\times 10^{-4} ) $                 &  (4.72)                  \\  \hline

$B^-\rightarrow \eta_cK^- $         &  $7.42\times 10^{-5}  $     &   -0.25   &  $7.42\times 10^{-5} $  &   -0.27    \\
						& $(5.54\times 10^{-5} ) $         &  (0.00 )      & $( 5.54\times 10^{-5} ) $     &  (0.00)    \\

$B^-\rightarrow D^0D^-_s $       &   $1.47\times 10^{-2} $   &   -0.21      &  $1.48\times 10^{-2} $     &   -0.23     \\
						&  ( $1.65\times 10^{-2} $)   & (-0.21 )     &  ($1.66\times 10^{-2} $)     &  (-0.23)    \\     \hline

$B^-\rightarrow \phi K^- $             & $6.21\times 10^{-6} $      &  0.74     &  $5.97\times 10^{-6} $         &   0.83   \\
						& $ (7.75\times 10^{-6} )$    & (0.58)   & $ (7.44\times 10^{-6} )$ & (0.65 ) \\
\hline       
\end{tabular}
\end{center}
\end{table}
\end{small}

\begin{small}
\begin{table}
\begin{center}
\caption{Branching  ratios and CP asymmetries  with $q^2=m^2_b/2 $   and  $N_c=2.4 $ ( entries in bracket correspond to the uncoupled case) }

\begin{tabular}{|c|lr|lr|}
\hline
&& \\
CKM  Matrix    &  $\rho=0.30, $ &  $    \eta=0.42  $    &   $ \rho=-0.20,  $   &   $   \eta=0.45 $   \\
\hline
Decay channel  &   BR                   &  $         a_{CP}(\%)       $  &     BR            &         $ a_{CP}(\%)       $     \\
\hline
$B^-\rightarrow \eta_c\pi^- $        &  $1.31\times 10^{-5}  $&               1.80   &  $1.32\times 10^{-5}  $ &   
1.90    \\
  	                                                  & $(1.29\times 10^{-5}) $  &             (0.00 )   &  $ (1.29\times 10^{-5} ) $      &  (0.00 ) \\

$B^-\rightarrow D^0D^- $              &  $3.38\times 10^{-4} $      & 3.74      & $3.06\times 10^{-4} $   &   4.45  \\
						&  $ (3.98\times 10^{-4}) $  & (3.80)   &  $(3.61\times 10^{-4} ) $                 &  (4.52)                  \\
\hline
$B^-\rightarrow \eta_cK^- $         &  $3.12\times 10^{-4}  $     &   -0.12   &  $3.11\times 10^{-4} $  &   -0.13    \\
						& $(3.08\times 10^{-4} ) $         &  (0.00 )      & $( 3.08\times 10^{-4} ) $     &  (0.00)    \\

$B^-\rightarrow D^0D^-_s $       &   $1.29\times 10^{-2} $   &   -0.19      &  $1.30\times 10^{-2} $     &   -0.20     \\
						&  ( $1.45\times 10^{-2} $)   & (-0.19 )     &  ($1.46\times 10^{-2} $)     &  (-0.21)    \\     
\hline
$B^-\rightarrow \phi K^- $             & $7.36\times 10^{-6} $      &  0.58     &  $7.06\times 10^{-6} $         &   0.65   \\
						& $ (9.14\times 10^{-6} )$    & (0.43)   & $ (8.75\times 10^{-6} )$ & (0.48 ) \\
\hline       
\end{tabular}
\end{center}
\end{table}
\end{small}

\newpage
\begin{flushleft}
{\Large\bf Figure Captions} \\
Fig.1:  Inelastic scattering $B^-\rightarrow D^{*0}D^-_s\rightarrow \phi K^- $ through $D^*_s  $  
exchange\\
Fig.2: Inelastic scattering $B^-\rightarrow D^0D^{*-}_s\rightarrow \phi K^- $ through $D^*_s $  
exchange\\

\newpage

\par
~~~~\\

\let\picnaturalsize=N
\def\picsize{3.0in}
\def\picfilename{paper.eps}
\ifx\nopictures Y\else{\ifx\epsfloaded Y\else\input epsf \fi
\let\epsfloaded=Y
\centerline{\ifx\picnaturalsize N\epsfxsize \picsize\fi \epsfbox{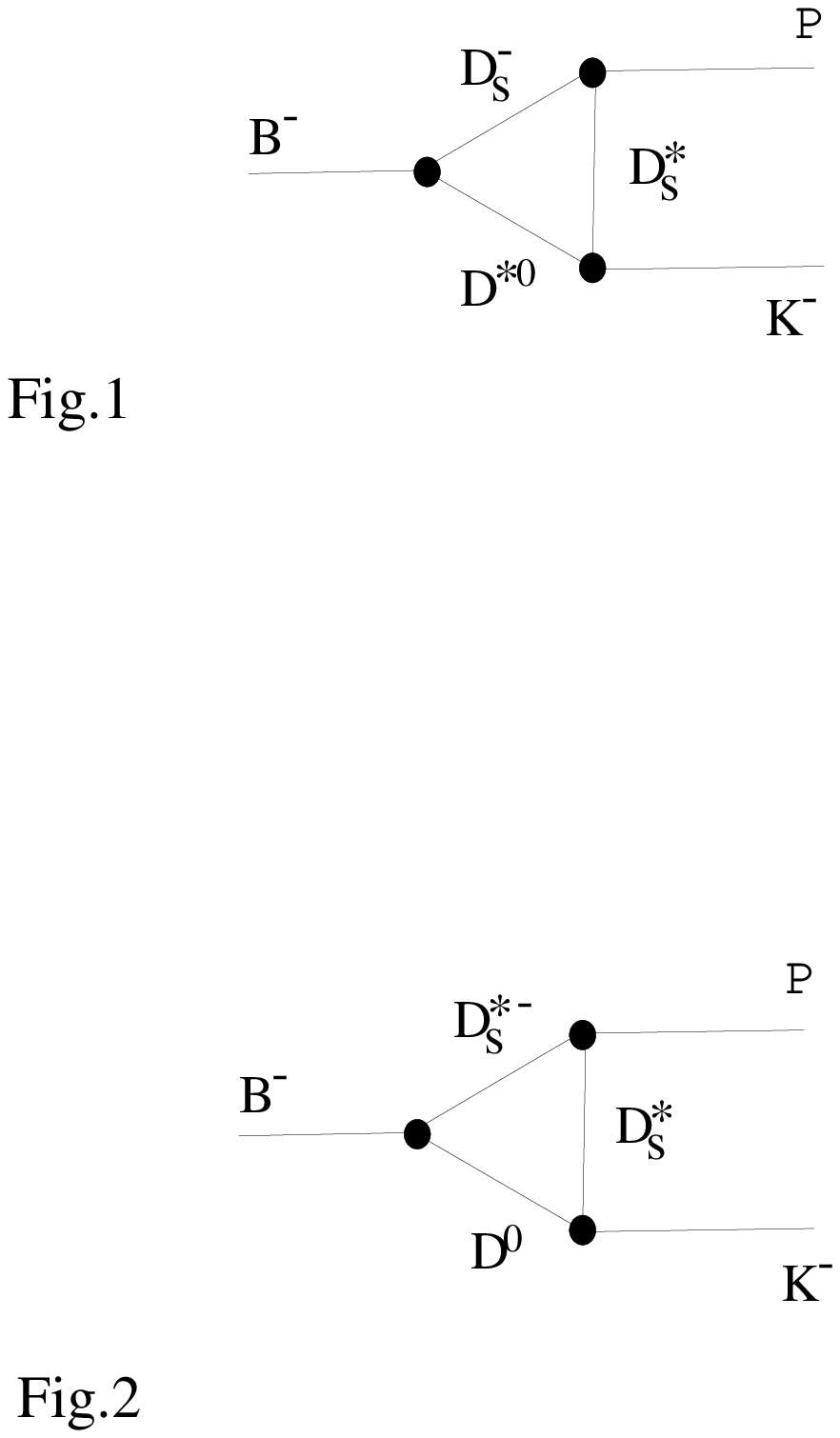}}}\fi

\end{flushleft}

\end{document}